\documentclass[fleqn,twoside]{article}
\usepackage{espcrc2}

\newcommand{\nl}{\nonumber\\}
\newcommand{\nn}{\nonumber}
\newcommand{\ds}{\displaystyle}
\newcommand{\lpar}{\left(}    
\newcommand{\rpar}{\right)}

\newcommand{\lcbr}{\left\{}
\newcommand{\rcbr}{\right\}}
\newcommand{\drii}[2]{\delta_{#1#2}}
\newcommand{\pmom}{p} 
\newcommand{\pone}{p_1}
\newcommand{\ptwo}{p_2}

\newcommand{\pmomi}[1]{p_{#1}}
\newcommand{\pmoms}{p^2}
\newcommand{\mf}{m_f}
\newcommand{\mz}{M_{_Z}}
\newcommand{\mws}{M^2_{_W}}
\newcommand{\mzs}{M^2_{_Z}}

\newcommand{\propf}[2]{{1\over{#1^2 + #2}}}
\newcommand{\Rxi}{R_{\xi}}
\newcommand{\gpar}{\xi}
\newcommand{\gparA}{\xi_{_A}}
\newcommand{\gparZ}{\xi_{_Z}}
\newcommand{\gpars}{\xi^2}
\newcommand{\gparAs}{\xi^2_{_A}}

\newcommand{\gparZs}{\xi^2_{_Z}}
\newcommand{\bq}{\begin{equation}}
\newcommand{\eq}{\end{equation}}
\newcommand{\bqa}{\begin{eqnarray}}
\newcommand{\eqa}{\end{eqnarray}}
\newcommand{\ba}[1]{\begin{array}{#1}}
\newcommand{\ea}{\end{array}}
\newcommand{\fep}{e^{+}}
\newcommand{\fem}{e^{-}}
\newcommand{\ft}{t}
\newcommand{\ab}{A}
\newcommand{\zb}{Z}
\newcommand{\wb}{W}
\newcommand{\wbpm}{W^{\pm}}
\newcommand{\hb}{H}
\newcommand{\hkn}{\phi^{0}}

\newcommand{\hkpm}{\phi^{\pm}}

\newcommand{\fpxpm}{X^{\pm}}
\newcommand{\fpyZ}{Y^{\ssZ}}
\newcommand{\fpyA}{Y^{\ssA}}
\newcommand{\ssA}{{\scriptscriptstyle{A}}}
\newcommand{\ssZ}{{\scriptscriptstyle{Z}}}
\newcommand{\sss}[1]{\scriptscriptstyle{#1}}
\newcommand{\ib}{i}
\newcommand{\gadi}[1]{\gamma_{#1}}

\newcommand{\gxd}{\gamma_6}
\newcommand{\gnd}{\gamma_7}
\newcommand{\qf}{Q_f}

\newcommand{\cow}{c_{_W}}
\newcommand{\siws}{s^2_{_W}}

\newcommand{\tcif}{I^{(3)}_{f}}

\newcommand{\AmS}{{\protect\the\textfont2
  A\kern-.1667em\lower.5ex\hbox{M}\kern-.125emS}}

\title{Project {\tt SANC} (former {\tt CalcPHEP}):
       Support of Analytic and Numeric calculations for experiments at Colliders}

\author{A. Andonov\address[MCSD]{Laboratory of Nuclear Problems, 
        Joint Institute for Nuclear Research, Dubna, Russia}
        \thanks{Supported by INTAS grant N$^{{o}}$ 00-00313},
        D. Bardin\addressmark$^*$,
        S. Bondarenko\addressmark$^*$,
        P. Christova\addressmark$^*$,
        L. Kalinovskaya\addressmark$^*$,
        G. Nanava\addressmark$^*$,\\
        G. Passarino\address{Department of Theoretical Physics, University of Torino, 
        INFN, Torino, Italy}\thanks{Supported by the European Union under contract 
        HPRN-CT-2000-00149 and by MIUR under contract 2001023713${}_{}$006.}}
       
\begin{document}

\begin{abstract}
The project, aimed at the theoretical support of experiments at modern and future 
accelerators --- TEVATRON, LHC, electron Linear Colliders (TESLA, NLC, CLIC) and muon
factories, is presented. Within this project a four-level computer system is being created,
which must automatically calculate, at the one-loop precision level the pseudo- 
and realistic observables (decay rates and event distributions) for more and more complicated 
processes of elementary particle interaction, using the principle of knowledge storing.

It was already used for a recalculation of the EW radiative corrections for Atomic Parity
Violation~\cite{Bardin:2001ii} and complete one-loop corrections for the process
$e^+ e^-\to t\bar{t}$~\cite{Bardin:2000kn,Andonov:2002rr,Andonov:2002xc}; for the latter an, 
agreement up to 11 digits with FeynArts and the other results is found. 
The version of {\tt SANC} that we describe here is capable of automatically computing 
the decay rates and the distributions for the decays $Z(H,W)\to f \bar{f}$ 
in the one-loop approximation.
\vspace{1pc}
\end{abstract}

\maketitle

\section{SANC PROJECT AND ITS ROOTS}
The main goal of this project is the creation of a  \linebreak
software product, accessible via Internet, for the  \linebreak
automatic computation of pseudo- and realistic      \linebreak
observables with a one-loop precision for various   \linebreak
processes of elementary particle interactions, such \linebreak
as: $1\to 2$, $1\to 3$, $2\to 2$, $1\to 4$, $2\to 3$, etc.~\cite{Bardin:2002gs}.

It has two roots: 1) Codes aimed at a theoretical support of HEP experiments, such as
{\tt TOPAZ0}~\cite{Montagna:1998kp},
{\tt ZFITTER}~\cite{Bardin:1999yd} and similar ones.\\
2) Numerous {\tt FORM2}-codes, written by the authors of ref.~\cite{Bardin:1999ak}
while they were working on it.

It is supposed that the main software products of the project should be the
Monte-Carlo event generators that are being created in collaboration with 
S.~Jadach and Z.~Was from INF (Krakow, Poland) 
and B.F.L.~Ward from University of Tennessee (Knoxville, USA).
On top of these generators, for some processes the semi-analytic codes 
are also provided, the latter being used both for the cross-checks of MC generators
and for fits of inclusive observables to the theory predictions.


\section{BASIC NOTIONS}

\subsection{The SM Lagran\-gian in the $\Rxi$ gauge}
All the calculations start from the SM Lagran\-gian in the $\Rxi$ gauge. It depends on 
25 input para\-meters (IPS), fields, and on three gauge parameters:
\bq
{\cal{L}}={\cal{L}}(\mbox{IPS: 25 parameters,~fields,}\,\gparA,\,\gparZ,\,\gpar),
\eq
where the fields are: fermions, vector bosons, physical Higgs field, $\hb$, and 
unphysical fields, $\hkn,\,\hkpm,\,\fpyA,\,\fpyZ,\,\fpxpm$~\cite{Bardin:1999ak}.
We give the example of Feynman Rules
for vector boson propagators:
\[
\ba{ll}
\ab:&\hspace*{-3mm}
{\ds{ \frac{1}{\pmoms}
\lcbr \drii{\mu}{\nu}+\lpar\gparAs-1\rpar\frac{\pmomi{\mu}\pmomi{\nu}}{\pmoms}\rcbr}}, \\[1mm]
\zb:&\hspace*{-3mm}
{\ds{\propf{\pmom}{\mzs}\lcbr\drii{\mu}{\nu}+\lpar\gparZs-1\rpar
\frac{\pmomi{\mu}\pmomi{\nu}}{\pmoms+\gparZs\mzs}\rcbr}},  \\[1mm]
{\wbpm}:&\hspace*{-3mm}
{\ds{\propf{\pmom}{\mws}\lcbr\drii{\mu}{\nu}+\lpar\gpars -1\rpar 
\frac{\pmomi{\mu}\pmomi{\nu}}{\pmoms+\gpars\mws}\rcbr}}.
\ea
\]
\subsection{Scalar functions and reduction}
At present, {\tt SANC} knows how to deal with up to third-rank tensorial 
reduction to the usual scalar functions: $A_0,\,B_0,\,C_0$ and $D_0$;
and to the auxiliary scalar functions: $a_0,\,b_0,\,c_0$ and $d_0$,
which are due to particular form of photonic propagator in $\Rxi$ gauge.
{\bf A new fortran library} for numerical calculations of all these functions but one
($D_0$) is created and thoroughly tested by means of comparison with the other codes.
For the $D_0$ function, we use an old coding of {\tt TOPAZ0}~\cite{Montagna:1998kp}.

\subsection{Amplitude basis, scalar form factors, helicity amplitudes}
{\tt SANC} computes the one-loop covariant amplitude (${\cal{A}}$)
of a process parametrized in a certain basis
by a certain number of scalar form factors (SFF).
Next, it computes helicity amplitudes (HA) in terms of SFF.  
As an example, we present three covariant amplitudes 
of the decays $B(Q)\to  f (p_1)\bar{f}(p_2)$, where $B=H,Z,W$:
\bqa
&&\hspace*{-6mm}
\hb\to f \bar{f}: {\cal{A}} \propto I{{\bf\cal{F}}_{\sss{S}}},
\\
&&\hspace*{-6mm}
\mbox{1 structure ({S-basis}), {1 SFF}, {1 HA}};
\nl[1mm]
&&\hspace*{-6mm}
\zb\to f \bar{f}: {\cal{A}}\propto\ib\gadi{\mu}\gxd{\bf {\cal{F}}_{\sss{L}}}
                                             +\ib\gadi{\mu}    {\bf {\cal{F}}_{\sss{Q}}}
                                +m_f D_{\mu}{\bf {\cal{F}}_{\sss{D}}},
\nl
&&\hspace*{-6mm}
\mbox{3 structures ({LQD-basis}), {3 SFF}, {3 HA}};
\nl[1mm]
&&\hspace*{-6mm}
\wb\to u \bar{d}: {\cal{A}}\propto\ib\gadi{\mu}\gxd{\bf{\cal{F}}_{\sss{L}}}
                                             +\ib\gadi{\mu}\gnd{\bf{\cal{F}}_{\sss{R}}}
\nl
&&\hspace*{16mm}
                            +m_u D_{\mu}\gxd{\bf{\cal{F}}_{\sss{LD}}}
                            +m_d D_{\mu}\gnd{\bf{\cal{F}}_{\sss{RD}}},
\nl
&&\hspace*{-6mm}
\mbox{4 structures ({LRD-basis}), {4 SFF}, {4 HA}},
\nn
\eqa
where $D_{\mu}=\lpar\pone-\ptwo\rpar_{\mu}$.
Besides SFF, the HA depend on kinematical factors and coupling constants. For instance,
for $\zb$ decay three HA are:
\bqa
{\bf A^{\sss{Z}}_{0\pm\pm}}&\hspace*{-2.5mm}=\hspace*{-2.5mm}&\frac{g\mf}{\cow}
\Biggl[a_f{\bf{\cal{F}}_{\sss{L}}}
 +\delta_f{\bf{\cal{F}}_{\sss{Q}}}
 +\frac{1}{2}a_f\beta_f^2\mzs{\bf{\cal{F}}_{\sss{D}}}
\Biggr],
\nl
{\bf A^{\sss{Z}}_{\pm\pm\mp}} &\hspace*{-2.5mm}=\hspace*{-2.5mm}& \frac{g\mz}{\sqrt{2}\cow}
\biggl[a_f(1\mp\beta_f){\bf{\cal{F}}_{\sss{L}}}+\delta_f{\bf{\cal{F}}_{\sss{Q}}}
\biggr],
\eqa
where $\beta^2_f = 1-4{\ds \frac{\mf^2}{\mzs}}\,,$ \\
and $\delta_f=v_f-a_f=-2\qf\siws,\;\; a_f=\tcif.$\\[1mm]
We note that the number of SFF and the number of independent non-zero HA always 
coincide. 
In general, a $2 f \to2 f $ process with four different external fermion masses 
is described by 16 SFF and 16 independent HA.
In case of $\fep\fem\to\ft\bar{t}$ process, 
one has six SFF and six independent HA if the electron mass is 
ignored~\cite{Bardin:2000kn,Andonov:2002rr,Andonov:2002xc}.

\section{STATUS OF THE PROJECT}
\subsection{Four levels of computer system} 
Here we explain what the {\tt SANC} system does at its four levels in the case of  
calculation of pseudo-observables (decay rates, event distributions) for 
the simplest decays: $H(Z,W)\to f_1\bar{f}_2$. 

{\bf Level 1:} the chain of calculations {\em 
from ${\cal{L}}_{\sss{SM}}$ to the ultraviolet-free helicity amplitudes} is realized.
(All codes are written in {\tt FORM3}~\cite{Vermaseren:2000nd}). 
It contains four sublevels, which calculate:
\vspace*{-2.3mm}
\begin{enumerate}
\item[$-$] the scalar form factors;
\vspace*{-2.8mm}
\item[$-$] the {\em soft} and {\em hard} photonic contributions to the decay rates
(analytic chain);
\vspace*{-2.8mm}
\item[$-$] the helicity amplitudes for basic processes;
\vspace*{-2.8mm}
\item[$-$] the helicity amplitudes for accompanying bremsstrahlung processes;
\vspace*{-2mm}
\end{enumerate}

{\bf Level 2:} an {\tt s2n.f} software generates automatically the {\tt fortran} codes for 
$\Gamma^{(1)}=\Gamma^{\rm Born}+\Gamma^{\rm Virt}+\Gamma^{\rm Soft}+\Gamma^{\rm Hard}$;

{\bf Level 3:} an infrared rearrangement (or exponentiation) procedure 
should work here (it is still at the stage of development);

{\bf Level 4:} a Monte Carlo event generator works out. For the time being,
we have here a `manually written' {\tt fortran} code. 

\subsection{Some keywords, characterizing {\tt SANC}} 
{\tt SANC} is an {\bf Internet-based} system 
(address {\it http://brg.jinr.ru/}) the use of which is supposed to be as simple as the usual 
surf on Internet.
It is a {\bf database-based} system. This means that there is
a storage of source codes written in seve\-ral languages, which talk to one another, being
placed into a homogeneous environment written in {\tt JAVA} (linker). 
It uses the {\bf precomputation or intermediate access principle} which means that: \\
1) All the relevant one-loop diagrams are precomputed and stored. This greatly saves 
the CPU. \\
2) It has several intermediate `entries' to bypass CPU consuming computations.
The user may also access its final product, a MC generator. 
From the other side, the full chain of calculation `from the Lagrangian to realistic 
distribution' can be worked out in real time upon the user request. This option is intended
to demonstrate the self-reproducibility of the full chain of calculation at any time.


\section{First versions of {\tt SANC}}
\vspace*{-1mm}
\begin{itemize}
\item {\bf v0.01, 3/01} realizes a part of analytic
calculations of level 1
for the decays $H$ $(Z,W)\to f_1\bar{f}_2$ (demonstration of viability);
\vspace*{-6.5mm}
\item several versions, {\bf v0.02c/d}, were tested on the road toward
realization of levels 2,4 for decays~$H(Z,W)\to f_1\bar{f}_2$;
\vspace*{-2.5mm}
\item the current version, {\bf v0.03} (summer'02), realizes the full chain of calculations,   
returns numbers and distributions for these decays at the one-loop level of precision
(demonstration of workability);
\vspace*{-2.5mm}
\item the next version, {\bf v0.10} (at work), should contain all {\tt FORM3} codes
of level 1 for the processes $2 f \to 2 f $ and the decays $F\to 3 f $.
\end{itemize}
\vspace*{-4mm}

\subsection{Comparison with other codes}
A lot of comparisons with the other codes were made.
For instance, we compared numbers with those of the {\tt ZFITTER} code
for all channels of $e^+ e^- \to f\bar f$ with light fermion masses
and found the following agreement:
within {\bf 8--9  digits} for the SFF; within {\bf 7--8  digits} for the  
one-loop differential cross-sections $d\sigma^{(1)}/d \cos\vartheta$;
within {\bf 6--7  digits} for the total cross section and $\sigma^{\sss FB}$.

Several comparisons were made for the $e^+ e^- \to t\bar t$ process.
First is an internal one, between two {\tt fortran} codes: {\tt s2n.f} and {\tt eeffLib}.
The latter is a `manually written' code aimed at benchmarking of {\tt s2n.f} software.
For the SFF and for the complete one-loop differential cross-section 
${d\sigma^{(1)}}/{d\cos\vartheta}$ the numbers agree within {\bf 12--13 digits}.
A comparison between {\tt s2n.f} and {\tt FeynArts} ~\cite{Hahn:2000jm}
for the one-loop cross-section {\em without soft photons} revealed an 
agreement within {\bf 11 digits}.
Finally, we compared {\tt s2n.f} numbers with those of the {\tt topfit} code
~\cite{Fleischer:2002rn,Fleischer:2002nn}
for the differential one-loop cross-section 
{\em with soft photons} and found an {\bf 8-digit} agreement.

\section{CONCLUDING REMARKS}

{\tt SANC} clearly is a long-term project. Its first phase is
realized in {\it demonstration version 0.10}. 

Its second phase assumes the creation of a complete software product,
accessible via an Internet-based environment, 
and realizing the full chain of calculations 
at the one-loop level of precision including 
{\em processes 2 $\to$ 3} and {\em decays 1 $\to$ 4}.


\begingroup\begin{thebibliography}{10}

\bibitem{Bardin:2001ii}
D.~Y. Bardin, P.~Christova, L.~Kalinovskaya and G.~Passarino, {\em Eur. Phys.
  J.} {\bf C22} (2001) 99--104.

\bibitem{Bardin:2000kn}
D.~Bardin, L.~Kalinovskaya and G.~Nanava, ``An electroweak library for the
calculation of EWRC to $e^+ e^- \to f  \bar{f}$ within the {\tt CalcPHEP} project'', 
CERN-TH/2001-308, {\tt hep-ph/0012080}.

\bibitem{Andonov:2002rr}
A.~Andonov, D.~Bardin, S.~Bondarenko, P.~Christova, L.~Kalinovskaya and G.~Nanava,
``Further study of the $e^+ e^- \to f \bar{f}$ process with the aid of {\tt CalcPHEP}
system'', 
CERN-TH/2002-068, {\tt hep-ph/0202112}.

\bibitem{Andonov:2002xc}
A.~Andonov, D.~Bardin, S.~Bondarenko, P.~Christova, L.~Kalinovskaya and G.~Nanava,\linebreak
``Update of one loop corrections for $e^+ e^- \to f  \bar{f}$, 
first run of {\tt CalcPHEP} system'', {\tt hep-ph/0207156}.

\bibitem{Bardin:2002gs}
D.~Bardin {\em et~al.}, 
``Project {\tt CalcPHEP}: Calculus for precision high energy physics'',
{\tt hep-ph/0202004}.

\bibitem{Montagna:1998kp}
G.~Montagna, O.~Nicrosini, F.~Piccinini and G.~Passarino, {\em Comput. Phys. Commun.} 
{\bf 117} (1999) 278.

\bibitem{Bardin:1999yd}
D.~Bardin, M.~Bilenky, P.~Christova, M.~Jack, L.~Kalinovskaya, A.~Olshevsky,
S.~Riemann and T.~Riemann, {\em Comput. Phys. Commun.} {\bf 133} (2001) 229-395.

\bibitem{Bardin:1999ak}
D.~Y. Bardin and G.~Passarino, {\em The standard model in the making: Precision
  study of the electroweak interactions}.
\newblock Oxford, UK: Clarendon (1999) 685 p.

\bibitem{Vermaseren:2000nd}
J.~A.~M. Vermaseren, ``New features of FORM'',
{{\tt math-ph/0010025}}.

\bibitem{Hahn:2000jm}
T.~Hahn, {\em Nucl. Phys. Proc. Suppl.} {\bf 89} (2000) 231--236.

\bibitem{Fleischer:2002rn}
J.~Fleischer, T.~Hahn, W.~Hollik, T.~Riemann, C.~Schappacher and A.~Wertenbach, 
``Complete electroweak one-loop radiative corrections to top-pair production at TESLA: 
-- a comparison'', {\tt hep-ph/0202109}.

\bibitem{Fleischer:2002nn}
J.~Fleischer, J.~Fujimoto, T.~Ishikawa, A.~Leike, T.~Riemann, Y.~Shimizu and A.~Wertenbach, 
``One-loop corrections to the process $e^+ e^- \to t  \bar{t}$
 including hard bremsstrahlung'',
{{\tt hep-ph/0203220}}.

\end{thebibliography}\endgroup

\end{document}